\documentclass{jetpl}
\usepackage{amsfonts,amssymb,amsmath,bm}
\usepackage{color}
\usepackage{graphicx}

\twocolumn

\lat

\title{Nonlinear generation of vorticity in thin smectic films}

\rtitle{Nonlinear generation of vorticity in thin smectic films}

\sodtitle{Nonlinear generation of vorticity in thin smectic films}

\author{V.\,M.\,Parfenyev\/\thanks{e-mail: parfenius@gmail.com}, S.\,S.\,Vergeles and V.\,V.\,Lebedev }

\rauthor{V.\,M.\,Parfenyev, S.\,S.\,Vergeles and V.\,V.\,Lebedev}

\sodauthor{Parfenyev, Vergeles, Lebedev}

\address{L.D.Landau Institute for Theoretical Physics RAS, 117940 Moscow, Russia}

\dates{11 December 2015}{*}

\abstract{We analyze a solenoidal motion in a vertically vibrated freely suspended thin smectic film. We demonstrate analytically that transverse oscillations of the film generate two-dimensional vortices in the plane of the film owing to hydrodynamic nonlinearity. An explicit expression for the vorticity of the in-plane film motion in terms of the film displacement is obtained. The air around the film is proven to play a crucial role, since it changes the dispersion relation of transverse oscillations and transmits viscous stresses to the film, modifying its bending motion. We propose possible experimental observations enabling to check our predictions.}


\begin{document}

\maketitle

{\bfseries 1. Introduction.} An interplay between vortical and wave motions is one of the most exciting and long-standing problem in hydrodynamics. We analyze this phenomenon for the case of freely suspended thin liquid films. On scales larger than the film thickness, such film can be considered as a two-dimensional system embedded into three-dimensional space. In comparison with traditional two-dimensional systems, such films possess an additional degree of freedom associated with their bending distortions. Therefore in addition to the usual in-plane hydrodynamic modes an extra mode can be exited in the freely suspended films associated just with the bending motion. The non-linear interaction of the bending motion with the in-plane hydrodynamic flows gives rise to generation of the solenoidal in-plane motion by the bending waves. We examine the phenomenon in our work.

One of widely known type of liquid films that can be freely suspended are soap films. They are considered as a model system providing an opportunity for testing two-dimensional hydrodynamic theory \cite{gharib1989}. However the detailed analysis has shown that the fluid motion in soap films is more complex and that a relation to motion described by the two-dimensional hydrodynamic equations is not straightforward \cite{couder1989}. The generation of a vortical motion in vibrating soap films has a long history. It was first observed by Taylor more than a century ago \cite{taylor1878}. Quantitative experiments were carried out at the end of the 20th century, see, e.g. \cite{airiau1986,afenchenko1998}, and a qualitative theoretical analysis is presented in the paper \cite{vega1998}. To our knowledge there is no quantitative theory of the phenomenon nowadays.

In our work, we consider freely suspended thin smectic films, see e.g. \cite{pieranski1993,Yablonskii1999}. Smectic liquid crystals are remarkable layered materials which can form a rich variety of structures. The simplest smectic structure is the smectic $A$ phase, which is solid-like in the direction perpendicular to the layers and fluid-like within the layers. The smectic $C$ and hexatic liquid crystals possess orientationally ordered layers. Any case, the smectic liquid crystal can be thought as a stack of fluid layers. Due to this layered structure thin films of the smectic liquid crystals consisted of a number of layers can be formed. Such films can be easily pulled from a reservoir of the substance. Due to the absence of the crystalline order inside the smectic layers such films can be treated as two-dimensional fluids. We develop a theory for the simplest case of the smectic $A$ films. However, it can be easily generalized to the smectic $C$ and hexatic films.

In our mind there are no reports about the vorticity generation in oscillating smectic films up to the date, while their transverse (bending) oscillations are well studied \cite{kraus1998, KL93}. Fortunately, dynamics of the smectic films is simpler than one of the soap films since there is no analog of Marangoni waves \cite{afenchenko1998} in the case. It allows us to construct a quantitative theory of the phenomenon. We consider and compare two cases, the film surrounded by vacuum and by air. It turns out that the air environment plays an important role in the film dynamics. The vortical fluid motion in the film appears due to nonlinear interaction of the bending waves, which is taken into account in the framework of a perturbation theory. Thus, our consideration is correct for sufficiently small amplitudes of transverse oscillations. We also discuss qualitatively the case of strong non-linearity.

{\bfseries 2. Film in vacuum.} Let us describe dynamical properties of a freely suspended thin film, pulled from the bulk smectic $A$ phase. Such film is an isotropic two-dimensional system, as we treat it on scales larger than the film thickness. We assume that in equilibrium the film is parallel to the $X-Y$ plane and that its bending distortions are characterized by the displacement $h(t,x,y)$ of the film in the $Z$-direction, i.e. the film shape is determined by the equation $z=h(t,x,y)$. The unit vector perpendicular to the film is ${\bm l}=(-\partial_x h,-\partial_y h, 1)/\sqrt{g}$, where $g=1+ (\nabla h)^2$ can be thought as the determinant of the film metric tensor.

The film state is described in terms of the film mass density $\rho(t,x,y)$ and of the film momentum density $\bm j (t,x,y)$. Both these quantities are two-dimensional densities of the film projection to the $X-Y$ plane. The dynamic equations for the film displacement $h$ and for the mass density $\rho$ are \cite{KL93}
 \begin{eqnarray}
 \rho\partial_t h
 =j_{z}-j_\alpha\partial_\alpha h,
 \qquad \partial_t \rho
 =-\partial_\alpha j_\alpha.
 \label{filmm1}
 \end{eqnarray}
Here and below Greek indices run over $x,y$. The first equation in (\ref{filmm1}) is the kinematic condition implying that the film moves with the velocity $\bm v =\bm j/\rho$, and the second equation is the mass conservation law.

The equations (\ref{filmm1}) should be supplemented by the equation for the momentum density $\bm j$ of the film. First we consider the film surrounded by vacuum. In the case the momentum is conserved, the corresponding equation reads \cite{KL93}
 \begin{equation}
 \partial_t j_i = -\partial_\alpha\left(v_\alpha j_i
 - \sqrt{g} \sigma \delta_{i\alpha}^{\perp}
 - \sqrt{g} \eta_{i\alpha \beta m}\partial_\beta v_{m}\right),
 \label{film3}
 \end{equation}
where Latin subscripts run over $x,y,z$ and $\delta^\perp_{ik}\equiv\delta_{ik}-l_i l_k$ stands for the projector to the film, and $\sigma$ is its surface tension. As above, all quantities are assumed to be functions of $t,x,y$. The viscosity  tensor can be written as \cite{KL93}
 \begin{eqnarray}
 \eta_{iklm}=(\zeta-\eta)
 \delta^{\perp}_{ik}\delta^{\perp }_{lm}
 +\eta(\delta ^{\perp }_{il}\delta^{\perp}_{km}
 +\delta ^{\perp }_{im}\delta ^{\perp }_{kl}) .
 \label{film5}
 \end{eqnarray}
Here $\eta>0$ and $\zeta>0$ are two-dimensional analogs of the three-dimensional first (shear) and second (bulk) viscosity coefficients. The equations (\ref{filmm1}-\ref{film5}) constitute a closed system of non-linear equations that describe the film dynamics.

We assume that the film bending distortion is weak, $|\nabla h| \ll 1$. All other quantities characterizing the film are assumed to deviate weakly from their equilibrium values as well. Then one can use the perturbation series in examining the film distortions. In the linear approximation there are two sound-like modes \cite{SM}. The first one is the bending sound associated with the bending distortions of the film. In the mode, each film element oscillates in the direction transverse to the film. The bending sound propagates with the velocity $c_b=\sqrt{\sigma_0/\rho_0}$, where $\rho_0$ and $\sigma_0$ are equilibrium values of the two-dimensional mass density $\rho$ and of the surface tension $\sigma$,  respectively. The second mode is the longitudinal sound, which is associated with fluctuations of the two-dimensional mass density $\rho$ and does not disturb the film shape. The sound propagates with the velocity $c_l=\sqrt{-\partial\sigma/\partial\rho}$. Further we assume that the longitudinal sound is not excited by the pumping force directly.

The linear equation for the bending sound reads
 \begin{equation}
 \partial_t^2 h +2 \hat\alpha \partial_t h
 -c_b^2 \nabla^2 h=0,
 \label{linear}
 \end{equation}
where the operator $\hat\alpha$ determines the sound attenuation. If the freely suspended film is surrounded by vacuum then the bending mode has an anomalously weak attenuation, the property is related to the rotational symmetry of the system \cite{KL93}. Particularly, there is no contribution to the attenuation $\hat\alpha$ caused by the film viscosity (\ref{film5}). There are some other contributions to the attenuation. First of all, there is a contribution of higher order in $\nabla$, proportional to $\nabla^4$. Second, there is a contribution related to the thermal fluctuations \cite{KL93}, it is caused by the non-linear interaction of the fluctuations. However, for films of the thickness of about hundred layers, we have in mind, the fluctuation contribution is relatively weak. In addition, there is a contribution to the bending mode attenuation $\hat\alpha$ related to processes in the meniscus. The contribution needs an additional investigation and it is beyond the scope of present work.

Our goal is to describe the in-plane vortical (solenoidal) motion generated by the bending motion owing to non-linear effects. The in-plane solenoidal motion is characterized by the vertical component of the vorticity $\varpi_z = \partial_x v_y -\partial_y v_x$. In the linear approximation $\varpi_z$ is zero, that is one should go beyond the linear approximation to find $\varpi_z$. We calculate the main nonlinear contribution to $\varpi_z$, which is of the second order in the film displacement $h$. The starting point of the calculation is the equation Eq.~(\ref{film3}), one should take its curl and project it to the $Z$-direction. Then one obtains an equation for $\varpi_z$:
 \begin{eqnarray}
 \rho_0 \partial_t \varpi_z-\eta\nabla^2 \varpi_z
 = \sigma_0 \epsilon_{\beta\gamma}
 \partial_{\beta} h \partial_{\gamma} \nabla^2 h
 \nonumber \\
 +\eta \epsilon_{\beta\gamma} \partial_\beta \partial_t
 (\partial_\gamma h \nabla^2 h),
 \label{varpiz}
 \end{eqnarray}
where $\epsilon_{\alpha\beta}$ is the unit two-dimensional antisymmetric tensor.

Let us analyze a steady contribution to the vorticity $\varpi_z$. Averaging over time the equation (\ref{varpiz}), one finds
 \begin{equation}
 \nabla^2 \varpi_z  = (\sigma_0/\eta) \epsilon_{\beta \gamma}
 \langle \partial_{\gamma} h \partial_{\beta} \nabla^2 h \rangle,
 \label{vort4}
 \end{equation}
where angle brackets designate time averaging. At the next step we use the linear equation for the bending sound (\ref{linear}) to obtain
 \begin{equation}
 \varpi_z  = 2(\rho_0/\eta) \epsilon_{\beta \gamma} \nabla^{-2}
 \langle \partial_{\gamma} h \partial_{\beta} \partial_t \hat\alpha h \rangle.
 \label{vvort4}
 \end{equation}
Let us stress that the expression (\ref{vvort4}) is proportional to the attenuation of the bending sound.

{\bfseries{3. Air Environment.}} Here we consider the case where the freely suspended smectic film is surrounded by air and analyze its influence to the film dynamics. As previously, we assume that some pumping force excites the bending oscillations. One can think about sound as the pumping source, the corresponding experimental technique is described in Ref. \cite{Uto97}. It is well known that the air around the soap film considerably changes its dynamics. Inspired by papers \cite{afenchenko1998, vega1998}, we consider an influence of air to the motion of the thin smectic film and, particular, to the mechanism of the vorticity generation.

We assume that the air sound velocity $c_a$ is much larger than $c_b$. It is reasonably to expect since $c_b^2=\sigma_0/\rho_0$ is inversely proportional to the number of the smectic layers in the film, and we have in mind the film consisting of about hundred layers. At the condition $c_a\gg c_b$ the air motion can be described in terms of the three-dimensional Navier-Stokes equation \cite{LL}, supplemented by the incompressibility condition $\mathrm{div}\; \bm v = 0$, where $\bm v$ is the air velocity.

The film separates two regions of space filled with air. In the presence of air, Eq.~(\ref{film3}) has to be modified, since the air influences the liquid motion in the film. The stress tensor in the air is
 \begin{equation}
 \sigma_{ik} = -p \delta_{ik} + \rho_a \nu_a (\partial_k v_i +\partial_i v_k),
 \label{stress}
 \end{equation}
where $\rho_a$ and $\nu_a$ are the air mass density and its kinematic viscosity coefficient, respectively, and $p$ is pressure.
Then the correct dynamic equation for the film is \cite{LL}
 \begin{equation}
 \label{air1}
 \begin{aligned}
 \partial_t j_i = -\partial_\alpha\left(v_\alpha j_i
 - \sqrt{g} \sigma \delta_{i\alpha}^{\perp}
 - \sqrt{g} \eta_{i\alpha \beta m}\partial_\beta v_{m}\right) &
 \\
 + \sqrt{g} (\sigma_{ik}^{II}-\sigma_{ik}^{I}) l_k &,
 \end{aligned}
 \end{equation}
where the unit vector $\bm l$, normal to the film, is pointed out from the region I ($z<h$) to the region II ($z>h$). Note that the velocity of air $\bm v$ is continuous in the whole space and it coincides with the film velocity at the film surface. The equation (\ref{air1}) can be treated as the boundary condition for the three-dimensional air motion.

The air motion around the film can be easily analyzed in the linear approximation. Exploiting the Navier-Stokes equation and the boundary condition (\ref{filmm1}), one obtains \cite{SM}
 \begin{eqnarray}
 \label{eq:vel1}
 \displaystyle v_{\alpha} = \mp \nu_a \frac{\hat{\kappa}(\hat{\kappa}
 + \hat{k})}{\hat{k}} \left(e^{\mp \hat{k}z} -  e^{\mp \hat{\kappa} z} \right) \partial_{\alpha} h,&&
 \\ \label{eq:vel2}
 \displaystyle v_z = \nu_a (\hat{\kappa} + \hat{k}) \left( \hat{\kappa} e^{\mp \hat{k}z}
 - \hat{k} e^{\mp \hat{\kappa} z} \right) h,&&
 \end{eqnarray}
where the upper/lower signs correspond to the regions II/I and we have introduced the following (non-local) operators $\hat{k}= (-\partial_x^2-\partial_y^2)^{1/2}, \, \hat{\kappa} = (\partial_t/\nu_a + \hat{k}^2)^{1/2}$. The first terms in Eqs.~(\ref{eq:vel1}, \ref{eq:vel2}) correspond to the potential part of the velocity, while the second terms belong to its solenoidal part. Further we assume that the air viscous term is weak in comparison with the frequency $\omega$ of the external force, that is
 \begin{equation}
 \gamma\equiv \sqrt{\nu_a k^2/ \omega} \ll 1,
 \label{gamma}
 \end{equation}
where $k$ is the wave vector of the bending mode excited in the film. Then the vortical (solenoidal) velocity is located in a much thinner layer of depth $1/\kappa \sim \gamma/|k|$ than the potential one, penetrating to the length $|k|^{-1}$.

Hereinafter we assume that the bending mode decay is supplied mainly by the air viscosity, that is correct if $|k|^{-1}$ is much larger than the film thickness. Then the bending dispersion law is \cite{SM}
 \begin{equation}
 \label{eq:dispersion}
 \omega^2 = \omega_0^2 \left( 1 - \frac{i \gamma}{\sqrt{2}} \Theta  \right),
 \quad \omega_0^2 = \frac{\sigma_0 |k|^2}{\rho_0 + 2 \rho_a/|k|},
 \end{equation}
where $\Theta =\left( 1 + \rho_0 |k|/2 \rho_a \right)^{-1}$. The term $2\rho_a/|k|$ corresponds to the associated mass of air involved into the bending motion of the film. The factor $1/|k|$ in $2 \rho_a/|k|$ stands for the penetration depth of the air potential velocity. The expression (\ref{eq:dispersion}) is obtained at the condition (\ref{gamma}).

Now we turn to the vorticity $\bm \varpi = \mathrm{curl}\, \bm v$. The bulk equation can be obtained by taking the curl of the Navier-Stokes equation \cite{LL}, it is
 \begin{equation}
 \label{eq:vorticity}
 \partial_t \bm \varpi = (\bm \varpi \nabla) \bm v
 - (\bm v \nabla) \bm \varpi + \nu_a \nabla^2 \bm \varpi.
 \end{equation}
In the linear approximation the vertical component of the vorticity, $\varpi_z$, is zero, it is generated due to nonlinearity. As before, we are interested in a contribution to $\varpi_z$ which is independent of time. Using Eq.~(\ref{eq:vorticity}) and averaging over time one finds
 \begin{equation}
 \label{eq:main}
 (\partial_z^2 - \hat{k}^2) \varpi_z =
 - \nu_a^{-1} \langle \varpi_{\alpha} \partial_{\alpha} v_z \rangle.
 \end{equation}
The term in the right hand side of the equation is a source for $\varpi_z$. The equation (\ref{eq:main}) has to be supplemented by the boundary condition (\ref{air1}). The internal film viscosity $\eta$ can be estimated as $\eta \sim \eta_{s} d$, where $d$ is the thickness of the film and $\eta_s$ is the dynamic viscosity coefficient of the bulk smectic. Thus the term with the film viscosity can be neglected for sufficiently thin films, $\rho_a \nu_a /\eta \gg |k|$. Then, using Eq.~(\ref{air1}), we obtain
 \begin{equation}
 \label{eq:BC}
 \left\langle (\partial_z \varpi_z)^{II} - (\partial_z \varpi_z)^{I} \right\rangle=0.
 \end{equation}
The condition (\ref{eq:BC}) should be posed at $z=0$. A solution of Eqs.~(\ref{eq:main}, \ref{eq:BC}) is
 \begin{equation}
 \begin{aligned}
 \label{eq:final}
 \varpi_z = \epsilon_{\alpha \beta} \left\langle \left( \dfrac{\hat{\kappa}}{\hat{k}}
 e^{\mp \hat{\kappa} z} \partial_{\alpha} h \right) e^{\mp \hat{k} z} \partial_{\beta} \partial_t h \right\rangle
 \\
 + (\nu_a \hat{k})^{-1} e^{\mp \hat{k} z} \epsilon_{\alpha \beta}
 \left\langle (\hat{k}^{-1} \partial_{\beta} \partial_t h ) \partial_{\alpha} \partial_t h \right\rangle.
 \end{aligned}
 \end{equation}
Similar to the case of the film surrounded by vacuum, the last term in the expression (\ref{eq:final}) is non-zero only if one takes into account the attenuation of the bending mode. For this reason, in calculations we have kept two first terms of the expansion in parameter $\gamma$. As a result we have obtained the first term in the expression (\ref{eq:final}). Details of the calculations can be found in \cite{SM}.

{\bfseries 4. Discussion.} We considered the case, where bending oscillations of the freely suspended smectic film are excited by an external pumping. We had in mind the situation where the pumping has frequencies in a narrow frequency range. Then some steady vortical (solenoidal) motions are excited in the film due to non-linear effects. We developed a quantitative theory enabling one to obtain an explicit expressions (\ref{vvort4}) or (\ref{eq:final}) for the vorticity of the film surrounded by vacuum and by air, respectively. Note that the expression (\ref{eq:final}) is correct when the air surrounding the film essentially influences the film dynamics. Particularly, the bending dispersion law is strongly modified in comparison with the sound one $\omega =  \sqrt{\sigma_0/\rho_0}\, k$, see Eq.~(\ref{eq:dispersion}).

Now we analyze the expression (\ref{vvort4}) and (\ref{eq:final}) to calculate the vertical steady vorticity $\varpi_z$ of the film for some specific case, where the film is bounded by a rectangular frame with dimensions $L_x$ and $L_y$ and a monochromatic pumping is applied to the film. Then the modes of the system are standing waves, which have zero displacement at the edges of the frame. Let us consider a superposition of two standing bending waves excited by the external pumping. Thus, we assume the film displacement to be
 \begin{equation}
 \begin{aligned}\label{eq:SW}
 h = H_1 \sin (k_x x) \sin (k_y y) \cos (\omega t) +
 \\
 + H_2 \sin (q_x x) \sin (q_y y) \cos (\omega t + \phi),
 \end{aligned}
 \end{equation}
where $k_x^2 + k_y^2 = q_x^2 + q_y^2 = k^2$, and the absolute value of the wave vector $k$ is determined by the frequency $\omega$ of the external force via the resonance condition $\omega= c_b k$ or (\ref{eq:dispersion}).

Taking into account the attenuation of the mode, and using Eq.~(\ref{vvort4}), we obtain the steady vorticity at the film embedded in vacuum
 \begin{equation}
 \label{eq:final02}
 \begin{aligned}
    \varpi_z    =    \dfrac{2 \alpha \rho_0}{\eta } \omega H_1 H_2
    \hat{k}^{-2} \sin \phi \times
    \\
    \Big[ k_y q_x  \sin (k_x x) \sin (q_y y) \cos(q_x x) \cos (k_y y) -
    \\
    - k_x q_y \cos (k_x x) \cos (q_y y) \sin(q_x x) \sin (k_y y) \Big].
 \end{aligned}
 \end{equation}
Above we have discussed different mechanisms contributed to the attenuation constant $\alpha$. Because of uncertainty in the relative effectiveness of the mechanisms, we do not know a dependence of the constant $\alpha$ on the wave vector $k$. The obtained vorticity $\varpi_z$ is proportional to the attenuation constant, therefore we conclude that the experimental study of the generated vorticity would provide a useful information concerning the attenuation of the transverse oscillations of the film surrounded by vacuum.

In the case of the film embedded in air, we can obtain the attenuation constant from the dispersion law~(\ref{eq:dispersion}), $\alpha =\omega \gamma \Theta / 2\sqrt{2}$, and using then Eq.~(\ref{eq:final}) we find the film vorticity
 \begin{equation}
 \label{eq:final2}
 \begin{aligned}
    \varpi_z (0)
    =    \dfrac{\omega H_1 H_2}{\gamma \sqrt{2}}
    \left(1 + \frac{\Theta}{2} |k| \hat{k}^{-1} \right) \sin \phi \times
    \\
    \Big[k_y q_x  \sin (k_x x) \sin (q_y y) \cos(q_x x) \cos (k_y y) -
    \\
    -k_x q_y \cos (k_x x) \cos (q_y y) \sin(q_x x) \sin (k_y y) \Big].
 \end{aligned}
 \end{equation}
Let us stress, that both answers, (\ref{eq:final02}) and (\ref{eq:final2}), are proportional to $\sin \phi$, where $\phi$ is a phase shift between the standing waves.

\begin{figure}[t]
\center{\includegraphics[width=1.0\columnwidth]{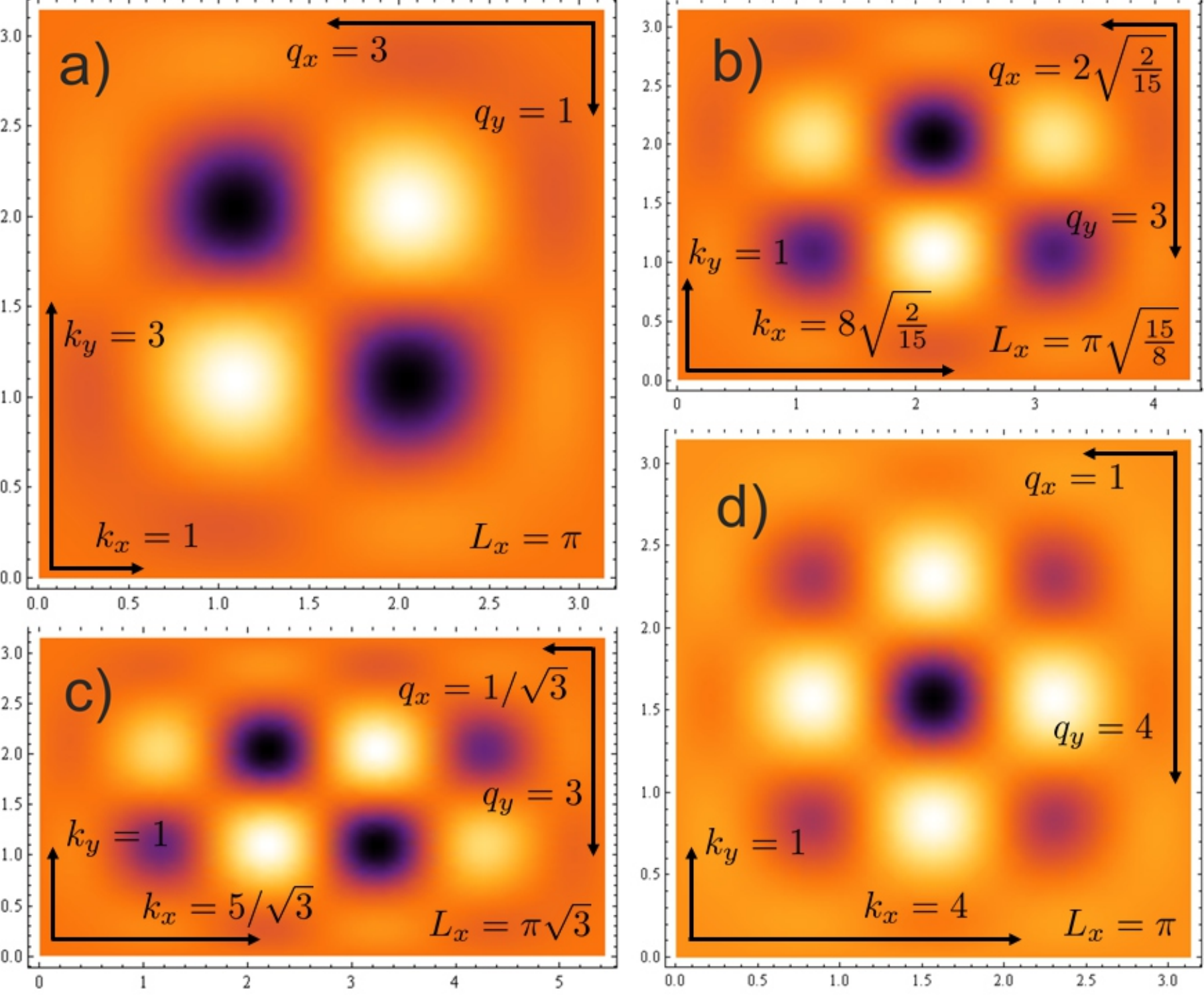}}
\caption{The spatial distribution of the steady vorticity $\varpi_z$
for two standing waves (\ref{eq:SW}). For all cases the second dimension of the frame $L_y=\pi$. The other parameters are shown in the figure.}
\label{pic:conf}
\end{figure}

Next we consider the film stretched in a nearly square frame. It this case $k_x=q_y$, $k_y=q_x$ and we deal with two degenerate modes. If the frame is perfectly square, then $\sin \phi = 0$ and there is no steady vorticity. However, slightly changing the aspect ratio of the frame one can produce the phase shift between the modes and the vertical vorticity becomes non-zero. When passing through the `point of compensation' the vorticity changes its sign. Some possible spatial vorticity distributions are presented in Fig.~\ref{pic:conf}. They were obtained numerically from the Eq.~(\ref{eq:final2}) in the limit $\Theta \rightarrow 1$. The vorticity found from Eq.~(\ref{eq:final02}) has similar spatial dependence.


In the end, let us formulate an applicability condition of our theory. It is correct when high-order nonlinear terms are smaller compared to the kept ones. In the case of the film surrounded by air, we have to estimate the nonlinear terms in Eq.~(\ref{eq:vorticity}) where the second-order terms for the velocity, $v^{(2)}$, have to be taken into account. It follows from Eq.~(\ref{eq:final}) that $v^{(2)}\sim\omega kh^2/\gamma$. Therefore the nonlinear terms with $v^{(2)}$ are small if $({\bm v}^{(2)}\nabla)\varpi_z \ll\nu\Delta\varpi_z$. Thus, we arrive at the condition $kh\ll \gamma^{3/2}$, that is stronger than the small tilt condition $kh\ll1$.

{\bfseries 5. Conclusion.} To summarize, we examined the nonlinear mechanism of the vorticity generation in thin freely suspended smectic $A$ films. We considered two cases, where the film is surrounded by vacuum and by air. In the first case the generated vorticity appears to be sensitive to the attenuation of the bending sound. We obtained the explicit formula (\ref{vvort4}) for the steady vorticity and speculated that an experimental study of the vorticity would provide a useful information about the anomalously weak attenuation of the bending sound. In the second case we found that the air considerably changes dynamics of the bending mode and affects the mechanism of the vorticity generation. We obtained the modified dispersion law (\ref{eq:dispersion}) and the explicit formula (\ref{eq:final}) for the vertical vorticity. We analyzed the formula and formulated some predictions which can be checked experimentally.

Though our final answers (\ref{vvort4}) and (\ref{eq:final}) are obtained for the case of monochromatic pumping, our theoretical scheme can be used for an arbitrary time dependence of pumping. Particularly, one can think about pumping containing two close frequencies. Then some beating effects in the generated vorticity are anticipated. The effects can be useful for experimental observations and to an experimental verification of our predictions.

Note that the applicability condition of our theory is $|\nabla h|\ll \gamma^{3/2}$, where $|\nabla h|$ is the film tilt and $\gamma$ is the small parameter of the theory (\ref{gamma}). If the tilt $|\nabla h|$ becomes larger than $\gamma^{3/2}$ then the applicability condition of our theory that is weakness of nonlinear effects in the film is violated. Then one expects a strong non-linearity of the solenoidal motions exited in the film. That could lead to formation of an analog of  the inverse cascade and to appearing of large coherent vortices in the film, like in Refs. \cite{Colm,Shats,14LBFKL}. Let us stress that the effect can be observed even for a weak tilt $|\nabla h|$.

We have developed our theory for the case of the smectic $A$ films that are isotropic. However, practically all the obtained results can be carried to the case of the orientationally ordered films of the smectics $C$ and of the hexatic smectics due to their weak anisotropy. Note that for such cases the vortical in-plane motion would generate some non-trivial orientational patterns that could help to investigate the motion.

We are grateful to S.~Yablonskii and E.~Kats for valuable discussions. This work was funded by Russian Science Foundation, Grant No. 14-22-00259.

\end{document}